\documentclass[showpacs,prl,aps,superscriptaddress,twocolumn]{revtex4-1}
\usepackage{amsmath,amsfonts,bm,graphicx,hyperref}

\begin{document}
\title{Electron Waiting Times in Mesoscopic Conductors}

\author{Mathias Albert}
\affiliation{D\'epartement de Physique Th\'eorique, Universit\'e de Gen\`eve, 1211 Gen\`eve, Switzerland}
\affiliation{Laboratoire de Physique des Solides, Universit\'e Paris Sud, 91405 Orsay, France}
\author{G\'eraldine Haack}
\author{Christian Flindt}
\author{Markus B\"uttiker}
\affiliation{D\'epartement de Physique Th\'eorique, Universit\'e de Gen\`eve, 1211 Gen\`eve, Switzerland}
\date{\today}

\begin{abstract}
Electron transport in mesoscopic conductors has traditionally involved investigations of the mean current and the fluctuations of the current. A complementary view on charge transport is provided by the distribution of waiting times between charge carriers, but a proper theoretical framework for coherent electronic systems has so far been lacking. Here we develop a quantum theory of electron waiting times in mesoscopic conductors expressed by a compact determinant formula. We illustrate our methodology by calculating the waiting time distribution for a quantum point contact and find a cross-over from  Wigner--Dyson statistics at full transmission to Poisson statistics close to pinch-off. Even when the low-frequency transport is noiseless, the electrons are not equally spaced in time due to their inherent wave nature. We discuss the implications for renewal theory in mesoscopic systems and point out several analogies with energy level statistics and random matrix theory.
\end{abstract}

\pacs{72.70.+m, 73.23.-b, 73.63.-b}


\maketitle

\textit{Introduction}.--- The distribution of waiting times between elementary physical events is an important tool to characterize and investigate temporal correlations and transport statistics of stochastic processes \cite{vankampen07}. Waiting time distributions (WTDs) play a significant role in various branches of science and technology, for instance in quantum optics \cite{Tannoudji86,Carmichael89} and single-molecule chemistry \cite{English06}. There, the dynamics is often Markovian and renewal theory applies such that the WTD fully defines a random walk in time and most statistical properties can be obtained from the WTD alone \cite{Cox62}. In contrast, for systems where renewal theory is not valid, the situation is more complex and various statistical quantities provide different information about a stochastic process. The dynamics is then similar to a random walk with memory effects. Investigations of WTDs have until now mostly dealt with either classical systems or Markovian quantum systems.

In electron transport, the waiting time between charge carriers has only received limited attention. One exception concerns weakly coupled nano-structures whose dynamics essentially is classical \cite{Scriefl05,Albert11}. In the opposite regime of fully coherent conductors, a theoretical framework for WTDs has so far been missing. The attention has instead been devoted to the full counting statistics (FCS) of transferred charges which for non-interacting electrons is elegantly expressed by the Levitov--Lesovik determinant formula \cite{Levitov93}. Typically, FCS concerns the limit of long times and important short-time physics may be lost (see however Ref.\ \cite{Ubbelohde12} and references therein). To describe short-time physics and correlations, WTDs are particularly useful and a theory of WTDs in mesoscopic conductors is thus of fundamental importance. Moreover, with the rapid progress in single-electron detection, there is hope that measurements of WTDs for a mesoscopic device could soon be within reach.

\begin{figure}
\includegraphics[width=0.92\columnwidth]{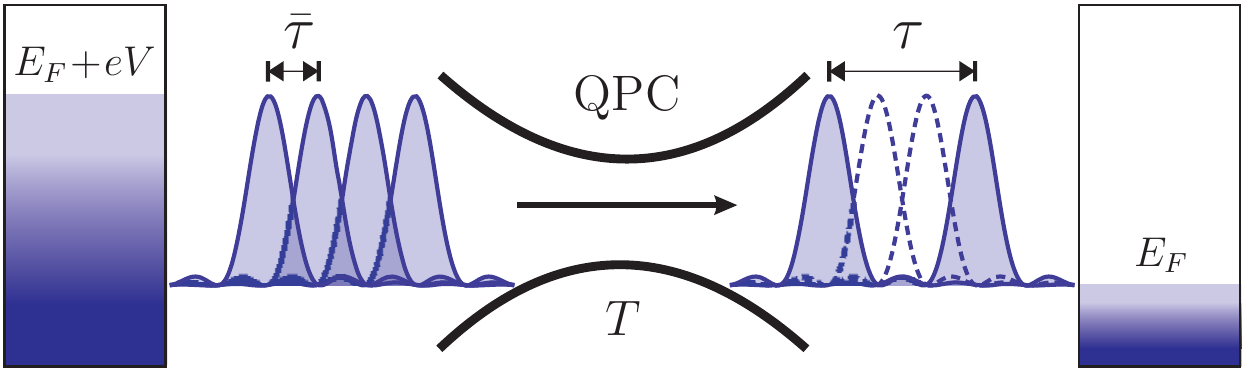}
\caption{(color online). Quantum point contact (QPC) connected to voltage-biased electrodes. The QPC transmission is denoted as $T$, the Fermi energy is $E_F$, and $V$ is the voltage applied to the source electrode. The average time separation of the in-coming electrons is $\bar{\tau}=h/eV$. The distribution of waiting times $\tau$ between transmitted electrons is determined by the many-body state of the in-coming electrons as well as the QPC which may cause electrons to reflect back. Reflected (missing) electrons are indicated by dashed lines. \label{Fig1}}
\end{figure}

In this Letter we develop a quantum theory of WTDs for mesoscopic conductors. The generic non-equilibrium system consists of electronic leads connected to a scatterer described by the scattering matrix $\mathcal{S}$. As a central result we formulate a compact determinant formula for the WTD in terms of the transmission amplitudes in $\mathcal{S}$. We employ a first-quantized many-body description which explicitly incorporates the fermionic statistics of the in-coming particles imposed by the Pauli exclusion principle \cite{Hassler08}. This is a generally applicable methodology. We illustrate it by calculating the WTD of a voltage-biased quantum point contact (QPC), Fig.\ \ref{Fig1}. For a fully open QPC, the low-frequency transport is known to be deterministic and regular \cite{Buttiker90}. Nevertheless, we find that the electrons are not equally spaced in time due to their inherent wave nature. Moreover, unlike the FCS, which is \emph{always} generalized binomial \cite{Hassler08,Abanov08}, the WTD exhibits a cross-over from Wigner--Dyson statistics for an open QPC to Poisson statistics close to pinch-off. These findings show how the WTD and the FCS provide separate, yet complementary, views on quantum transport in mesoscopic conductors. In general, renewal theory does not apply for mesoscopic conductors and no simple relations exist between the WTD and the long-time FCS. We conclude by pointing out a number of appealing analogies between WTDs and energy level statistics in random matrix theory \cite{Metha91}.

\textit{Formalism.}--- The distribution of waiting times $\tau$ between charge carriers is denoted as $\mathcal{W}(\tau)$. It can be expressed in terms of the idle time probability $\Pi(\tau)$: Given a random time $t_0$, the probability that \emph{no} electrons are detected in the time interval $[t_0,t_0+\tau]$  with $\tau \geq 0$ is denoted as $\Pi(\tau)$. For stationary processes $\Pi(\tau)$ does not depend on $t_0$. The WTD can be written
\begin{equation}
 \label{eq:itp2wtd}
  \mathcal{W}(\tau) = \langle \tau \rangle  \frac{d^2}{d\tau^2}{\Pi}(\tau).
\end{equation}
Here $\langle \tau \rangle$ is the mean waiting time which ensures the proper normalization $\int_0^\infty d\tau\mathcal{W}(\tau)=1$, since $\Pi(0)=1$ by definition and $\Pi(\tau)$ decays to zero at large times. This important relation can be derived by picking a random time $t_0$ and assuming that the \emph{last} prior detection of an electron occurred at the earlier time $t_e\leq t_0$. The idle time probability is then $\Pi(\tau)=\int_{-\infty}^{t_0}dt_e[1-\int_{t_e}^{t_0+\tau}ds \mathcal{W}(s-t_e)]/\langle \tau\rangle$, where the square brackets contain the probability that no electrons are detected in the time interval $[t_e,t_0+\tau]$. We also integrate over all possible times for the last electron to be detected, $-\infty\leq t_e\leq t_0$, using that detection events are uniformly distributed in time with weight $1/\langle\tau\rangle$. Finally, a suitable change of integration variables yields $\Pi(\tau)=\int^{\infty}_{\tau}du\int_{u}^{\infty}dv \mathcal{W}(v)/\langle \tau\rangle$ which explicitly shows that the idle time probability is independent of $t_0$ and immediately leads us to Eq.\ (\ref{eq:itp2wtd}). Importantly, in deriving Eq.\ (\ref{eq:itp2wtd}) we have made no Markov assumption nor relied on any renewal property.

Our next task is to evaluate the idle time probability for a generic mesoscopic conductor described by the scattering matrix $\mathcal{S}$. To this end we make use of the first-quantized many-body formalism developed by Hassler \emph{et al.}~\cite{Hassler08}. Our methodology applies to arbitrary geometries of the electronic leads and the scatterer, but to keep the discussion simple we consider a one-dimensional problem with a single in-coming and out-going quantum channel connected to the scatterer. The electrons are non-interacting and spinless and the temperature is zero. An applied voltage bias $V$ brings the system out of equilibrium such that in-coming electrons on the left side of the scatterer occupy states in the energy window $I_V=[E_F,E_F+eV]$ above the Fermi level $E_F$.  The scattering states take the form $\varphi_k(x)=e^{ikx}+r_ke^{-ikx}$, $x<0$, and  $\varphi_k(x)=t_ke^{ikx}$, $x>x_s>0$, where $r_k$ and $t_k$ are the reflection and transmission amplitudes in $\mathcal{S}$ and $k$ is the momentum. We need not specify the scattering states in the scattering region $[0,x_s]$.

We work close to the Fermi level, where the dispersion relation $E_k=\hbar v_F k$ is linear in $k$ and all wave components propagate with the Fermi velocity $v_F$. Next, we split the energy window $I_V$ into $N$ intervals of size $eV/N$, where $N$ is the number of incoming particles, and associate to the $m$'th energy interval a time-dependent wave function $\phi_m(x,t)=\langle x|\phi_m(t)\rangle=\int_{I_m}dk e^{-iv_Fkt}\varphi_k(x)/\sqrt{2\pi\kappa}$. Here  $I_m=\kappa [m-1,m]$ is the integration interval of width $\kappa= eV/N\hbar v_F$. All single-particle states are filled and the $N$-particle many-body state is given by the Slater determinant $|\Psi_S^{(N)}(t)\rangle$ of the states $|\phi_m(t)\rangle$, $m=1,\ldots,N$.

The detection of individual particles to the right of the scatterer, at $x_0>x_s$, is effected by the single-particle projection operator $\mathcal{Q}_{\tau}$. Its expectation value with respect to a single-particle state is the probability of detecting the given particle in the time interval $[t_0,t_0+\tau]$ or equivalently, given the linear dispersion, to detect the particle in the spatial range $I_\tau =[x_0,x_0+v_F\tau]$. We can therefore write $\mathcal{Q}_{\tau}=\int_{I_\tau} dx|x\rangle\!\langle x|$ \cite{Hassler08}. This is a projective measurement. The idle time probability for the Slater determinant is then $\Pi(\tau)=\langle \Psi_S^{(N)}(\tau)|\bigotimes_{i=1}^{N}(1-\mathcal{Q}_{\tau})|\Psi_S^{(N)}(\tau)\rangle$. The expectation value of such a product of single-particle operators with respect to a Slater determinant can itself be written as a determinant and we thereby obtain
\begin{equation}
\label{eq:itp}
\Pi(\tau)=\det(1-\mathbf{Q}_\tau).
\end{equation}
The \emph{single-particle} matrix elements of $\mathbf{Q}_\tau$ are $[\bold{Q}_\tau]_{m,n} = \langle\phi_m(\tau)|\mathcal{Q}_\tau|\phi_n(\tau)\rangle= \int_{I_m} dk' \int_{I_n}  dk  t^*_{k'} t_k   K_\tau (k-k') /2\pi\kappa$ with the kernel $K_\tau(q) = 2e^{- i q v_F \tau/2} \sin(q v_F \tau/2)/q$. Finally, to obtain a stationary process we take the limit $N\to\infty$ for which $\kappa\rightarrow 0$. We then have $[\bold{Q}_{\tau}]_{m,n} \simeq \kappa t^*_{\kappa m} t_{\kappa n} K_\tau(\kappa(n-m))/2\pi$ for $m,n=1,\ldots,N$. Equations (\ref{eq:itp2wtd}-\ref{eq:itp}) constitute the central results of this section as they allow us to calculate the WTD for an arbitrary voltage-biased scatterer.

\begin{figure*}
\includegraphics[width=0.85\textwidth]{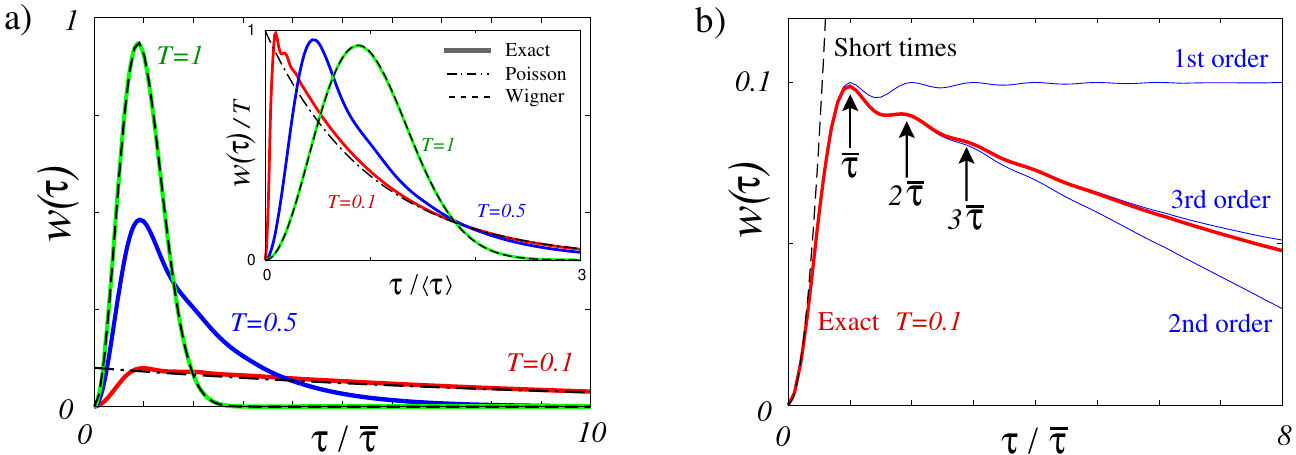}
\caption{(color online). Electron waiting times for a quantum point contact (QPC). {\bf a)} Waiting time distributions (WTDs) for different transmission probabilities $T$ of the QPC. The waiting time $\tau$ is given in units of $\bar{\tau}=h/eV$ or $\langle\tau\rangle =\bar{\tau}/T$ (inset). For a fully open QPC, $T=1$, the WTD is well approximated by a Wigner--Dyson distribution, Eq.\ (\ref{wigner}). Close to pinch-off, $T\simeq 0$, the WTD approaches Poisson statistics given by the exponential distribution $\mathcal{W}(\tau)\simeq e^{-\tau/\langle\tau\rangle}/\langle\tau\rangle$. {\bf b)} WTD in the low-transmission regime, $T\simeq 0.1$. The short-time behavior given by Eq.\ (\ref{eq:shorttime}) is indicated with a dashed line. Thin lines show expansions of the WTD to first, second, and third order in $T$. Arrows indicate small oscillations with period $\bar{\tau}$.  \label{fig:wtd}}
\end{figure*}

\textit{Quantum point contact.}--- We illustrate our method by calculating the WTD for a QPC with energy independent transmission $|t_{\kappa m}|^2=T$. Figure \ref{fig:wtd}a displays WTDs calculated for different transmission probabilities $T$. For a fully open QPC, $T=1$, the WTD reflects only the fermionic statistics and correlations in the many-body state. In particular, $\mathcal{W}(0)=0$, since two electrons cannot occupy the same state. This suppression is similar to the Fermi-hole in the density-density correlation function of a free electron gas \cite{Landau59}.  The fermionic correlations also force the WTD to decay strongly beyond a few mean waiting times, where it essentially vanishes. The mean waiting time is $\langle \tau\rangle = 1/\mathrm{Tr}(\mathbf{\dot{Q}}_{\tau=0})$ with $\mathbf{\dot{Q}}_\tau\equiv  \partial_\tau \mathbf{Q}_\tau$. It easily follows that $\mathrm{Tr}(\mathbf{\dot{Q}}_{\tau=0})=\mathrm{Tr}(\mathbf{\dot{Q}}_\tau)=(eV/h)\sum_{m=1}^N |t_{\kappa m}|^2/N=\langle I\rangle/e$ is the average (particle) current. For the QPC, we find $\langle \tau\rangle = \bar{\tau}/T$ with $\bar{\tau}=h/eV$. This is in line with the common wave-packet picture that electrons emitted from a constant-voltage source into a quantum channel on average are separated by the fundamental time scale $h/eV$ \cite{Landauer87}. However, the electrons are not equally spaced in time due to their inherent wave nature which introduces fluctuations in the waiting time. Interestingly, for $T=1$ the WTD is well approximated by the Wigner--Dyson distribution \cite{Metha91}
\begin{equation}\label{wigner}
\mathcal{W}_{\mathrm{WD}}(\tau) = \frac{32\tau^2}{\pi^2 \bar{\tau}^2} e^{-4\tau^2/\pi\bar{\tau}^2},\,\, \bar{\tau}=\frac{h}{eV}.
\end{equation}
This can be understood using analogies with energy level statistics and random matrix theory as we discuss below.

As the transmission $T$ is reduced below unity, the WTD widens as electrons may now reflect back on the QPC, allowing for longer waiting times between transmitted charges, Fig.\ \ref{fig:wtd}a. Close to pinch-off, $T\simeq 0$, the mean waiting time becomes much longer than $\bar{\tau}$ and the WTD approaches an exponential distribution corresponding to Poisson statistics involving rare and uncorrelated tunneling events. In this regime, the WTD is mainly determined by the scatterer, although the complete suppression at $\tau=0$ persists. In fact, for arbitrary transmission, we can expand the WTD to lowest order in $\tau$ as
\begin{equation}
\mathcal{W}(\tau)\simeq \frac{\pi^2}{3}\frac{T}{\bar{\tau}}  \left(\frac{\tau}{\bar{\tau}}\right)^2,\,\, \tau\ll\bar{\tau}.
\label{eq:shorttime}
\end{equation}
The Pauli exclusion principle manifests itself in a $\tau^2$-dependence of the WTD for short waiting times and a complete suppression at $\tau=0$. Equation~(\ref{eq:shorttime}) is indicated in Fig.\ \ref{fig:wtd}b and is in good agreement with the WTD for arbitrary transmissions $T$ as we have checked.

The WTD exhibits a cross-over from Wigner--Dyson statistics to Poisson statistics with decreasing transmission. The intermediate regime displays small oscillations with period $\bar{\tau}$ due to the quasi-regular train of incoming electrons superimposed on an exponential decay, Fig.~\ref{fig:wtd}b. Physically, the wave packets have a large overlap leading to small-amplitude oscillations. This situation resembles the density-density correlations of a liquid whose particles have a large degree of freedom and where the structure of the many-body state decays fast with the inter-particle distance. In contrast, for a solid-like system, whose constituents are placed in a nearly perfect ``crystal'', long-range correlations lead to well-isolated periodic peaks in the WTD. This behavior occurs in periodically driven classical systems \cite{Albert11} and is also expected for nearly non-overlapping quantum states produced by coherent single-electron sources \cite{Feve07}.

To further understand the low transmission regime, we expand the idle time probability in $T$. To this end we rewrite it as $\Pi(\tau)=\det(1-\mathbf{Q}_\tau)=\exp{[\mathrm{Tr}(\log\{1-\mathbf{Q}_\tau\})]}$. Since $\mathbf{Q}_\tau$ is proportional to $T$, we can expand the idle time probability as $\Pi(\tau)= 1-\mathrm{Tr}(\mathbf{Q}_\tau)+[\mathrm{Tr}(\mathbf{Q}_\tau)^2-\mathrm{Tr}(\mathbf{Q}^2_\tau)]/2+\mathcal{O}(T^3)$. The second derivative with respect to time is readily performed, allowing us to express the WTD as a series in $T$. Figure \ref{fig:wtd}b illustrates how the series improves as more terms are included. In particular, to fully account for the long-time behavior of the WTD high-order terms must be included. To lowest order in $T$ we find $\mathcal{W}(\tau)=\langle I\rangle g^{(2)}(\tau)/e+\mathcal{O}(T^2)$, where $g^{(2)}(\tau)=1-\sin^2(\pi t/\bar{\tau})/(\pi t/\bar{\tau})^2$ is the two-point correlation function \cite{Metha91}. This is consistent with renewal theory \cite{Carmichael89}. High-order terms, however, cannot be obtained from renewal theory. The break-down of renewal theory becomes evident by considering the cumulants of the WTD.

\begin{figure}
\includegraphics[width=0.95\linewidth]{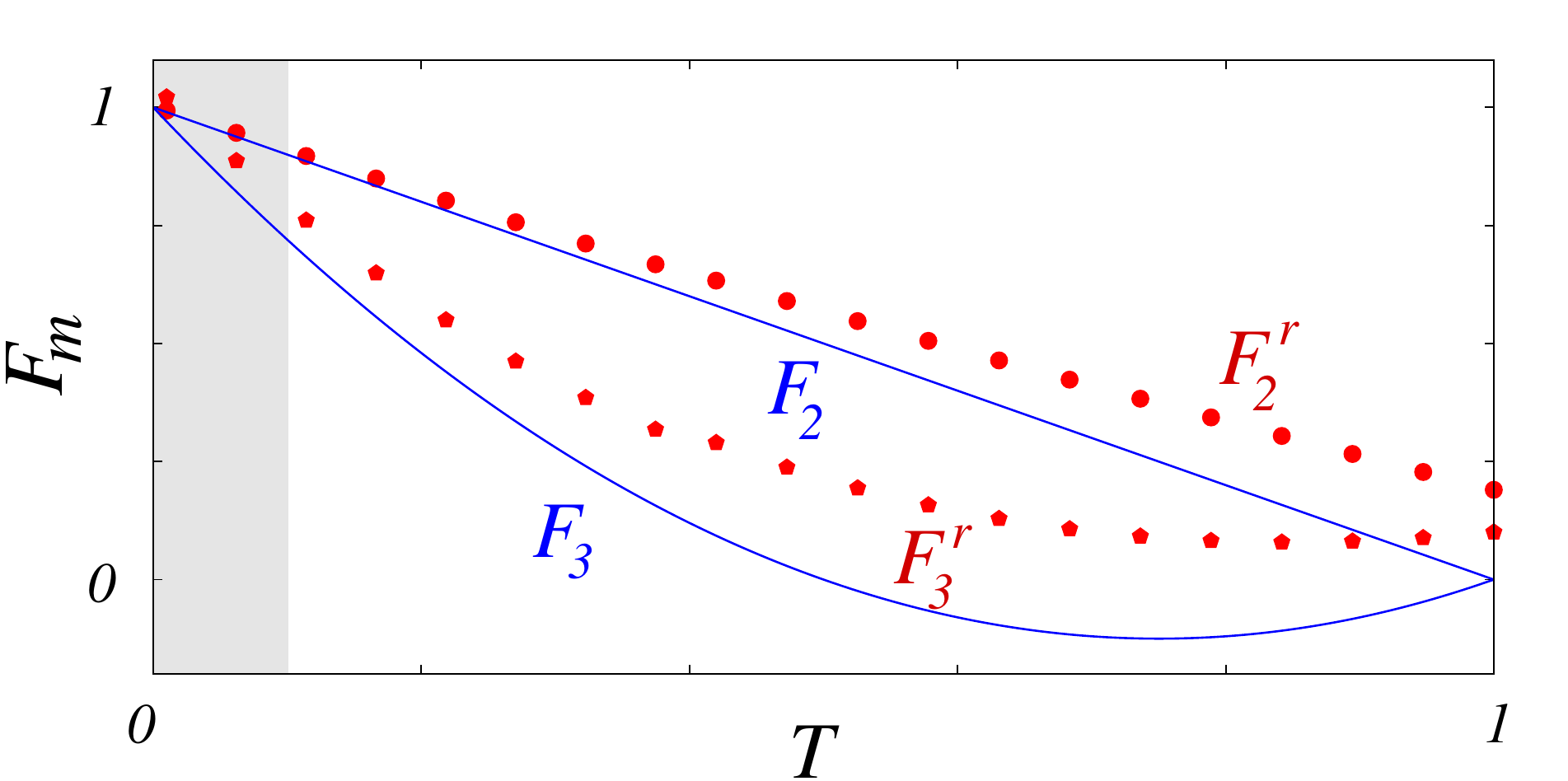}
\caption{(color online). Fano factors as functions of QPC transmission $T$. The exact results $F_2=1-T$ and $F_3=(1-T)(1-2T)$ are shown together with predictions based on renewal theory $F^{r}_2= \langle\!\langle \tau^2 \rangle\!\rangle/\langle\!\langle \tau \rangle\!\rangle^2$ and $F_3^{r} = 3 \langle\!\langle \tau^2 \rangle\!\rangle^2/\langle\!\langle \tau \rangle\!\rangle^4- \langle\!\langle \tau^3 \rangle\!\rangle/\langle\!\langle \tau \rangle\!\rangle^3$, where $\langle\!\langle \tau^m \rangle\!\rangle$ are the cumulants of the waiting time $\tau$. Renewal theory only holds in the low-transmission regime (shaded area), where $F_m\simeq F_m^r$.
 \label{Fig3}}
\end{figure}

\textit{Moments \& cumulants.}--- The first three cumulants of the WTD are the mean $\langle\!\langle \tau\rangle\!\rangle = \langle \tau\rangle$, the variance $\langle\!\langle \tau^2\rangle\!\rangle = \langle (\tau-\langle \tau\rangle)^2\rangle$, and the skewness  $\langle\!\langle \tau^3\rangle\!\rangle = \langle (\tau-\langle \tau\rangle)^3\rangle$. The cumulants of the FCS are similarly denoted as $\langle\!\langle n^m \rangle\!\rangle$, $m=1,2,3,\ldots$, where $n$ is the number of transferred charges. The corresponding Fano factors are $F_m \equiv \langle\!\langle n^m \rangle\!\rangle/\langle\!\langle n \rangle\!\rangle$, which for a QPC are well-known and read $F_2=1-T$ and $F_3=(1-T)(1-2T)$ \cite{Levitov93}. Importantly, for systems where renewal theory applies, the FCS can be directly related to the WTD and the Fano factors read $F^{r}_2= \langle\!\langle \tau^2 \rangle\!\rangle/\langle\!\langle \tau \rangle\!\rangle^2$ and $F_3^{r} = 3 \langle\!\langle \tau^2 \rangle\!\rangle^2/\langle\!\langle \tau \rangle\!\rangle^4- \langle\!\langle \tau^3 \rangle\!\rangle/\langle\!\langle \tau \rangle\!\rangle^3$ in terms of the waiting time $\tau$ \cite{Albert11,Budini11}. These relations provide us with a direct test of renewal theory. For a fully open QPC, the second Fano factor is zero, corresponding to regular transport without zero-frequency fluctuations. However, the width (or variance) of the WTD is finite, Fig.\ \ref{fig:wtd}a, explicitly demonstrating the break-down of renewal theory due to the fermionic statistics of the in-coming electrons. In Fig.\ \ref{Fig3} we examine the validity of renewal theory as a function of the QPC transmission. Only close to pinch-off ($T\simeq0$), where the transport is poissonian, we find $F_m\simeq F_m^r$ in agreement with renewal theory. In general, we expect that renewal theory for mesoscopic conductors is only valid in the low-transmission regime where tunneling events are rare and nearly uncorrelated.

\textit{Random matrix theory.}--- There are several analogies between WTDs and level spacing distributions in spectral statistics \cite{Metha91}. The latter are useful to discriminate complex systems according to their symmetry class or underlying classical dynamics and they may carry signatures of quantum chaos according to the Bohigas--Giannoni--Schmidt conjecture \cite{Bohigas84}. For a fully open QPC the analogy is due to the equivalence between the ground state of one-dimensional fermions and the joint probability distribution of eigenvalues of random matrices \cite{Len64,Calogero69,Metha91}. In particular, the canonical ensemble of random matrices, labeled by their symmetry parameter $\beta$, can be mapped onto the Calogero--Sutherland model of interacting one-dimensional fermions with coupling constant proportional to $\beta-2$ \cite{Calogero69}. Free fermions correspond to $\beta=2$ if we replace the $N$ coordinates by the $N$ eigenvalues of a random matrix from the gaussian unitary ensemble. As a direct consequence, all spatial correlation functions are given by the energy correlation functions in random matrix theory. Additionally, since the dispersion relation is linear, the system is invariant under Galilean transformation and the spatial correlations are identical to the temporal correlations. Finally, for an ensemble of random hermitian matrices of rank $N$ with $N\to \infty$, the Wigner--Dyson surmise is a good approximation of the level spacing distribution and it also agrees well with the WTD for a fully open QPC as we found in Fig.\ \ref{fig:wtd}a.

\textit{Conclusions}.--- We have presented a quantum theory of electron waiting time distributions (WTDs) in mesoscopic conductors expressed by a compact determinant formula. The WTD of a quantum point contact (QPC) exhibits a cross-over from Wigner--Dyson statistics for a fully open QPC to Poisson statistics close to pinch-off. We explicitly demonstrated the break-down of renewal theory for mesoscopic conductors. Finally, we discussed analogies between WTDs and level spacing distributions in spectral statistics.  Open questions to address in future work concern the influence of spin and finite-temperature effects as well as the WTDs of more complex scatterers.

\textit{Acknowledgments}.--- We thank  O.\ Bohigas, T.\ Brandes, P.\  Degiovanni, F.\ Hassler, and P.\ Jacquod for instructive discussions. The work was supported by Swiss NSF, MaNEP, NanoCTM, and NCCR QSIT.

\end{document}